\documentclass{PoS}

\usepackage{macros_static}
\usepackage{amsmath}
\usepackage{tikz}
\usepackage{wrapfig}
\usepackage{dcolumn,booktabs,colortbl}

\newcommand{\beq}{\begin{equation}}
\newcommand{\eeq}{\end{equation}}

\title{B $\to$ $\pi$ form factor with 2 flavours of $O(a)$ improved Wilson quarks}

\ShortTitle{B $\to$ $\pi$ form factor with $N_{\rm f}=2$ $O(a)$ improved Wilson quarks}

\author{Felix Bahr, \speaker{Fabio Bernardoni}, Alberto Ramos, Hubert Simma, Rainer Sommer\\%
       DESY, Zeuthen\\
       E-mail: \email{fabio.bernardoni@desy.de,felix.tobias.bahr@desy.de,\\
       alberto.ramos@desy.de,hubert.simma@desy.de,\\
       rainer.sommer@desy.de}}

\author{John Bulava\\
        CERN, Geneva\\
        E-mail: \email{john.bulava@cern.ch}}

\abstract{The determinations of $|V_{\rm ub}|$ from the exclusive
branching ratios of $B\to \tau \nu$ and $B \to \pi l \nu$ 
tend to show a tension at the level of $3\sigma$ \cite{Beringer:1900zz}. On the theoretical side they depend on the lattice computation 
of the hadronic matrix elements $f_{\rm B}$ and the $B\to \pi$  form factor 
$f_+(q^2)$. To understand the tension, improved precision and a careful analysis of the systematics involved are necessary. Working towards this goal, we present preliminary lattice results of the ALPHA collaboration for 
the $B\to \pi$  form factor $f_+(q^2)$ with $N_{\rm f}=2$ flavours of $O(a)$-improved 
Wilson fermions. Our computation uses HQET in the static limit, pion masses ranging 
down to $\sim250$ MeV, large volumes with $m_\pi L >4$, three lattice spacings, 
and non-perturbative renormalization. We describe the techniques adopted to reduce the statistical noise (stochastic all-to-all with full time dilution) and the contamination from excited states (smearing for the B and the pion). We estimate the size of the chiral and continuum extrapolations. We discuss the impact our result could have to clarify the above mentioned discrepancy in the determination of $|V_{\rm ub}|$.\vspace{0.9cm}
\begin{flushright}
DESY 12-152\\
SFB/CPP-12-67\\
CERN-PH-TH/2012-267
\end{flushright}
}

\FullConference{The 30 International Symposium on Lattice Field Theory - Lattice 2012,\\
		June 24-29, 2012\\
		Cairns, Australia}

\begin{document}

\section{Introduction}

The precise determination of the CKM matrix elements is a key for testing 
the Standard Model. 
Violations of CKM unitary or discrepancies between independent determinations 
of the same matrix element can provide hints for New Physics.
At the time when we started our work, a tension at the level of $3\sigma$  
between two exclusive determinations of $|V_{\rm ub}|$ existed, as reported e.g.
in the PDG of 2012. 
These determinations use the branching ratio (BR) for the processes $B\to \pi l 
\nu$ and $B\to \tau \nu$ from experiment combined with the form factor $f_+(q^2)$ 
and the B decay constant $f_{\rm B}$, respectively, from the lattice. 
Also an inclusive determination, based on a perturbative expansion in 
$\alpha_s$ and an expansion in $1/m_b$, is possible. 
The results reported by the PDG 2012, computed before ICHEP 2012, can be summarized as follows \cite{Beringer:1900zz}:
\begin{eqnarray}
|V_{\rm ub}| &=& 0.00323(31) \qquad B\to \pi l \nu \nonumber \\
|V_{\rm ub}| &=& 0.00510(47) \qquad B\to \tau \nu  \\
|V_{\rm ub}| &=& 0.00441(34) \qquad \mbox{incl.} \nonumber 
\end{eqnarray}
At ICHEP 2012 the Belle collaboration reported a new result for BR$(B\to \tau \nu)$ \cite{Yook:2012} based on a new set of data, obtained with a more sophisticated tagging of the B. This result, taken alone, would yield a 
value for $|V_{\rm ub}|$ that is consistent with the exclusive determination from $B\to \pi$. However, more data and a careful inspection of the systematics involved are needed to draw more definitive conclusions.\\
While the experimental precision in the differential decay rate for $B \to \pi l \nu$ has by now reached good precision, $B \to \tau \nu$ events are more difficult to reconstruct and an error of the order of $20 \%$ on the total branching ratio has to be expected. The situation on the theoretical side is the opposite: the lattice computation of a form factor is more challenging 
than a decay constant.
\\
The results for the determination of $f_{\rm B}$ by the ALPHA collaboration have been
presented in a poster at this conference \cite{Bernardoni:2012}. The aim of 
the work described here is the determination of $f_+(q^2)$ in the same setup, 
which uses the same CLS configurations. 
In particular in our final results we will use fully non-perturbative 
renormalization and matching.
At this conference we have presented the progress reached so far, putting some emphasis on the techniques employed to deal with the sizable finite-$T$ effects and with the signal-to-noise ratio problems.

\section{General Setup}
\begin{wrapfigure}{r}{0.5 \textwidth}
\begin{tikzpicture}[ scale=.7]
\vspace{-4cm}
\draw[fill=black]  (2,6) circle (0.2);
\draw[fill=black]  (6,6) circle (0.2);
\draw[fill=black] (10,6) circle (0.2);
\path[draw,line width=1pt]
 (2,6) -- (6,6);
\path[draw,line width=1pt]
(6,6.1) -- (10,6.1);
\path[draw,line width=1pt]
 (6,5.9) -- (10,5.9);
\path[draw,line width=1pt]
 (2,6) .. controls(6,8) .. (10,6);
\node  at (2,5.8) [below] {$\pi$};
\node  at (6,5.8) [below] {$V^\mu$};
\node  at (10,5.8) [below] {$B$};
\node  at (4,5.6) [below] {$t_\pi$};
\node  at (8,5.6) [below] {$t_{\rm B}$};
\end{tikzpicture}
\caption{Quark line diagram for the three point function.}
\label{quarkline}
\end{wrapfigure}
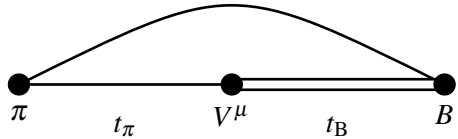

Neglecting the lepton masses, the SM prediction for the $B\to \pi l \nu$ decay rate is:
\beq
  \frac{\textnormal{d}\Gamma}{\textnormal{d}q^2} = \frac{G_\textnormal{F}^2 |V_{\rm ub}|^2}{24\pi ^3 } |p_\pi|^3  \left|f_+(q^2)\right|^2   
\qquad  
\eeq
where $q^\mu=p_{\rm B}^\mu-p_\pi^\mu$ and
the form factor $f_+(q^2)$ is defined through the Lorentz decomposition of the matrix element:
\beq
\langle \pi(p_\pi) | V^\mu | B(p_{\rm B} ) \rangle = f_+(q^2)\left[ p_{\rm B}^\mu + p_\pi^\mu-\frac{m_{\rm B}^2-m_\pi^2}{q^2}q^\mu \right]+f_0(q^2)\frac{m_{\rm B}^2-m_\pi^2}{q^2}q^\mu \,.
\eeq
To extract this matrix element from lattice simulations, we consider the ratio:
\beq
\label{ratio}
R(t_\pi,t_{\rm B})\equiv\frac{ 
\sum\limits _{\vec{x}_\pi, \vec{x}_{\rm B}}e^{-i\vec{p}  \cdot \vec{x}_\pi}
\langle   P_{ll}(t_\pi+t_{\rm B},\vec{x}_\pi)  V^\mu (t_{\rm B},\vec{x_{\rm B}})  P_{hl}(0)  \rangle}
{\sqrt{\sum\limits_{\vec{x}_\pi}e^{-i\vec{p}  \cdot \vec{x}_\pi}\langle  P_{ll}(x_\pi) P_{ll}(0)  \rangle  \sum\limits_{\vec{x}_{\rm B}} \langle  P_{hl}(x_{\rm B})   P_{hl}(0)  \rangle}}\,,
\eeq

\beq
\langle \pi(p_\pi) | V^\mu | B(p_{\rm B} ) \rangle =  \lim_{T,\,t_{\rm B},\,t_\pi\to \infty} R(t_\pi,t_{\rm B})e^{E_\pi t_\pi/2}e^{m_{\rm B} t_{\rm B}/2}
\eeq
where $P_{ll}$ and $ P_{hl}$ are interpolating operators for the $\pi$ and the $B$ meson, respectively. Our conventions for $t_{\rm B}$ and $t_\pi$ are illustrated in the quark line diagram, in Fig.~\ref{quarkline}.\\
We treat the $b$ quark in HQET, where a different Lorentz decomposition of $\langle \pi(p_\pi) | V^\mu | B(p_{\rm B} ) \rangle$ in the $B$ rest frame is more convenient:
\begin{eqnarray}
\langle \pi | V^0 |B\rangle &=& \sqrt{2 m_{\rm B}} f_\parallel\,, \nonumber \\
\langle \pi | V^k |B\rangle &=& \sqrt{2 m_{\rm B}} p_\pi^k f_\bot\,. \nonumber
\end{eqnarray}
In the static limit $f_+$ is proportional to $f_\bot$:
\beq
\label{stat}
f_+ = \frac{\sqrt{m_B}}{\sqrt{2}}f_\bot \,.
\eeq
\begin{wraptable}{r}{0.5\textwidth}
\begin{center}
\begin{tabular}{@{\extracolsep{0.2cm}}ccccc}
\toprule
id & $L/a$ & $a$ [fm] &  $m_\pi$[MeV] & $m_\pi L$  \\
\midrule
A2  & $32$  &  0.0755   &$630$  & $7.7$ \\ 
\midrule  
E4  & $32$  &  0.0658   &$ 580$ & $6.2$ \\
E5 &  $32$  &&$ 420$ & $4.7$ \\
F6  & $48$  &           &$ 310$ & $5.0$ \\
\midrule
O7  &$64$   &  0.0486   &$270$  & $4.2$ \\
\bottomrule
\end{tabular}
\end{center}
\caption{CLS ensembles used in this exploratory study.}
\label{quarkCLS}
\end{wraptable}
We have computed the ratio eq.~\eqref{ratio} on some of the CLS ensembles, as listed in Table~\ref{quarkCLS}, with different pion masses and three lattice 
spacings\cite{lambda:nf2}. These simulations use two $O(a)$-improved Wilson 
quarks in the sea and relatively fine lattice spacings.
Boundary conditions are periodic and the time extent is $T=2L$.
Finite volume effects are expected to contribute less than $1\%$ 
to the global systematic error, since $m_\pi L >4$ on all ensembles. \\
As the $b$ quark cannot be treated relativistically with present lattice spacings,
we employ HQET. This is an expansion of QCD in powers of $1/m_b$ 
and can be used to describe processes for which the relevant momentum scale
in the rest frame of the $B$-meson, in our case $|\vec{p_\pi}|$, 
satisfies $|\vec{p_\pi}|\ll m_b$. The 
main advantage of this approach is that HQET is renormalizable, so that the 
continuum limit can be taken and errors coming from this extrapolation can be 
estimated. Power divergences have to be removed non-perturbatively,
and of course also the renormalization of the vector current should be
done with non-perturbative precision.
We here use the results of \cite{DellaMorte:2006sv}, where first the
renormalization group invariant (RGI) renormalization factor was computed
non-perturbatively. Then the matching factors only contain the physical $m_b$-dependence through the ratio $M_b/\Lambda_{\overline{MS}}$ where $M_b$ is the RGI b-quark mass. The necessary matching factor, $C_V$ has been computed at 3-loops in continuum perturbation theory \cite{Chetyrkin:2003vi}. For the operator, which enters in the computation of the form factor 
$f_+(q^2)$ in the static limit, the renormalization and matching 
reads:
\begin{align}
V_k^{QCD}  &= C_V(M_b/\Lambda_{\overline{MS}})
Z_{A,RGI}^{stat}(g_0)V_k^{stat}+O(1/M_b)  \,.
\end{align} 
However, the accuracy of the perturbative expansion for B-physics has been questioned in \cite{Sommer:2010ic}. 
We use this perturbative (3-loop) matching only in the present
exploratory work, but will employ non-perturbative matching similar to~\cite{Blossier:2012qu}
in our future analysis.
We will then also include $O(a)$-improvement for the vector current
operator, while in the present setup we expect discretisation effects 
of $O(a \alpha_s)$.

\section{Techniques to reduce contamination(s) from
  excited states and finite $T$}

\begin{wrapfigure}{r}{0.5\textwidth}
\begin{center}
\includegraphics[width=10cm]{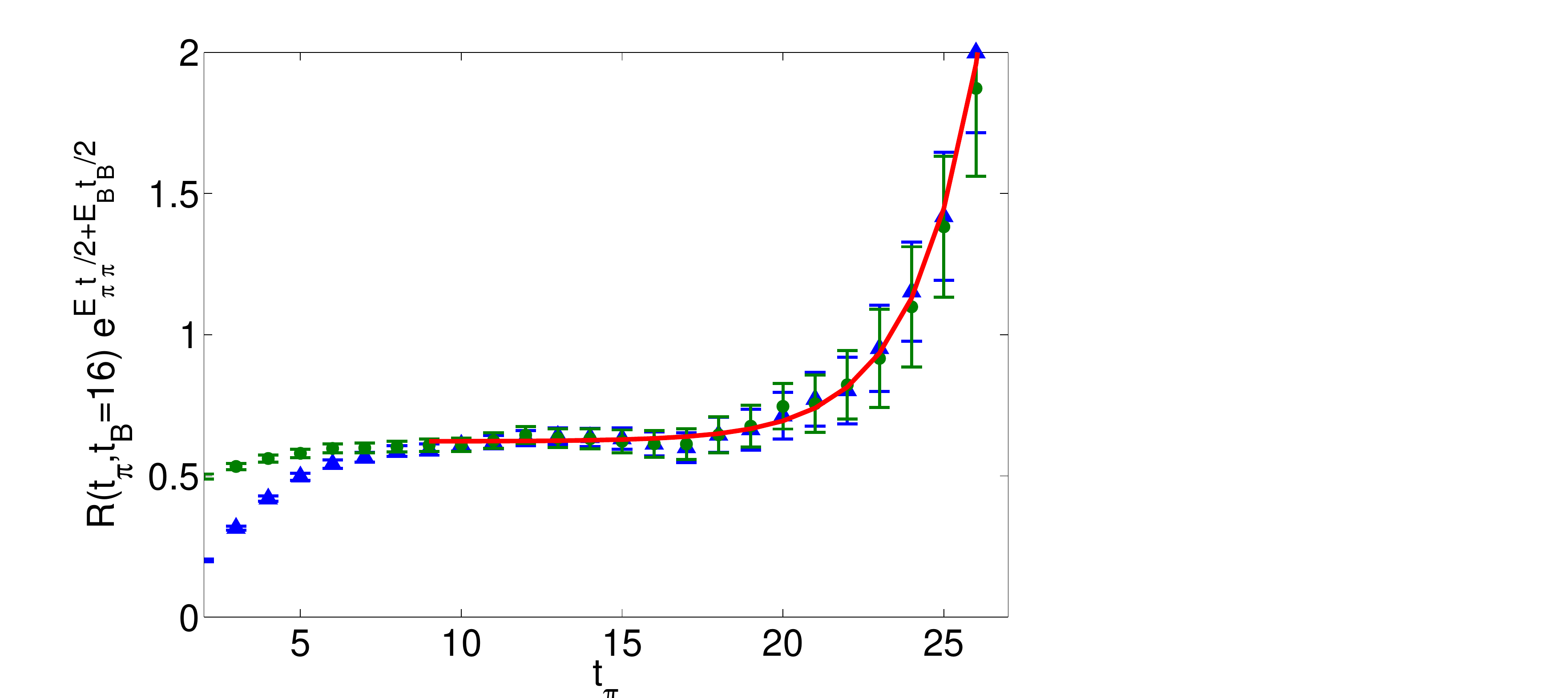} 
\end{center}
\caption{$t_\pi$ dependence for the considered ratio eq.~\protect \eqref{ratio}, for E5 lattice (see Table~\protect \ref{quarkCLS}). In these curves the interpolating operator for the B meson is always smeared. Circles and triangles correspond respectively to a local
and a smeared one for the pion. The red curve is obtained by fitting the finite $T$ effects with eq.~\protect \eqref{ratiofit}.}
\label{smearing}
\end{wrapfigure}
Significant contributions to the 3-point function in eq.~\eqref{ratio} 
may arise from a pion which propagates ``around the world'' in the 
$T$-periodic time direction of the lattice. 
Apart from a free pion state, also a $\pi$-$B$ state, which we denote by $[\pi B]$, 
contributes to these finite-$T$ effects.
This can be understood by considering the transfer matrix representation for
eq.~\eqref{ratio} at suitable time separations such that the contamination 
from other excited states is negligible. This is equivalent to the requirements
\beq
\Delta_\pi t_\pi\gg 1\ ,\quad  \Delta_\pi (T-t_\pi)\gg 1 \ ,\quad \Delta_{\rm B} t_{\rm B}\gg 1 
\,,
\eeq
where $\Delta_\pi$ and $\Delta_{\rm B}$ are the energy shifts between the ground 
state and the first excited state in the $\pi$- and in the $B$-channel, respectively. 
We then obtain:
\beq
\label{finiteT}
R(t_\pi,t_{\rm B})e^{E_\pi t_\pi/2}e^{M_{\rm B} t_{\rm B}/2} \rightarrow 
  \frac{
   \langle \pi | V^\mu | B \rangle+
 \langle 0             | V^\mu |  [\pi B]\rangle e^{-E_\pi(T-2t_\pi)-\Delta_{[\pi B]} t_{\rm B}}}
 {\sqrt{1+e^{-E_\pi (T-2t_\pi)}}} 
\eeq
where:
\beq
\Delta_{[\pi B]}=E_{[\pi B]} - E_\pi - E_{\rm B} \,.
\eeq
The finite-$T$ effect is hence proportional 
to the matrix element $\langle 0 | V^\mu |  [\pi B]\rangle$. 
If we assume (also based on numerical evidence) that $\langle 0| V^\mu| [\pi B]\rangle$
and $\langle \pi | V^\mu | B\rangle$ have comparable magnitude and that 
%
%
$\Delta_{[\pi B]} \approx 0$,
we conclude that finite-$T$ effects dominate for $t_\pi>T/2$. The region where one expects to observe an approximate plateau is limited to $t_\pi<T/2-k/E_\pi$ where $k$ depends on $\langle 0| V^\mu |  [\pi B]\rangle/\langle \pi | V^\mu | B\rangle$. It would be optimal to consider lattices with $T>2L$ or with other boundary conditions to 
compute these 3-point functions \cite{Luscher:2011kk}. Here, we just employ gaussian smeared wave
functions both for the B and the $\pi$, to allow the extraction of the matrix
element $\langle \pi | V^\mu | B \rangle $ at relatively
small time separations $t_\pi$ and $t_{\rm B}$.\\
Using smeared pion wave functions also helps to improve the 
signal-to-noise ratio which is of the order $e^{-(E_\pi-m_\pi)t_\pi}$ 
for the pion propagation. The effect of the smearing is shown in 
Fig.~\ref{smearing}. In the future we will also consider the possibility to
use non-spherical smearing for non-zero momentum pion wave functions \cite{DellaMorte:2012xc}.

\section{Results}

To extract the form factor on each ensemble we have tried two different methods. 
The first is a linear two-parameter fit at a fixed $t_{\rm B}$
\beq
\frac{
 A+B_{t_{\rm B}} e^{-E_\pi(T-2t_\pi)}}
 {\sqrt{1+e^{-E_\pi (T-2t_\pi)}}}
 \label{ratiofit}
\eeq
The value of $t_{\rm B}$ needs to be chosen small enough that the signal-to-noise ratio is satisfactory
but large enough that the excited states effects can be neglected.
The good description of the data (see red curve in Fig.~\eqref{smearing}) 
which we achieve with this fit ansatz indicates that the finite-$T$ effects 
are properly taken into account.\\
\begin{figure}
\begin{center}
\begin{tabular}{cc}
\includegraphics[width=11cm]{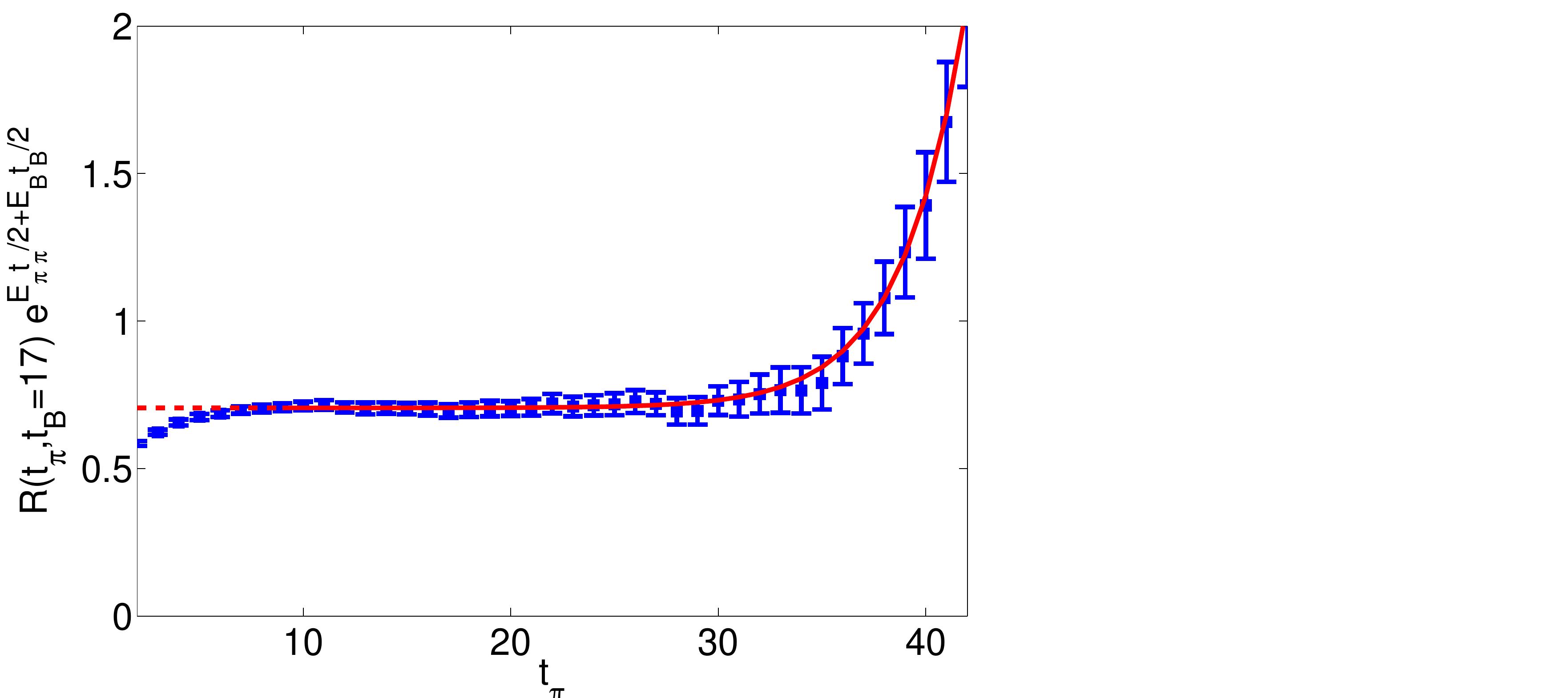}
&
\hspace{-4cm}\includegraphics[width=11cm]{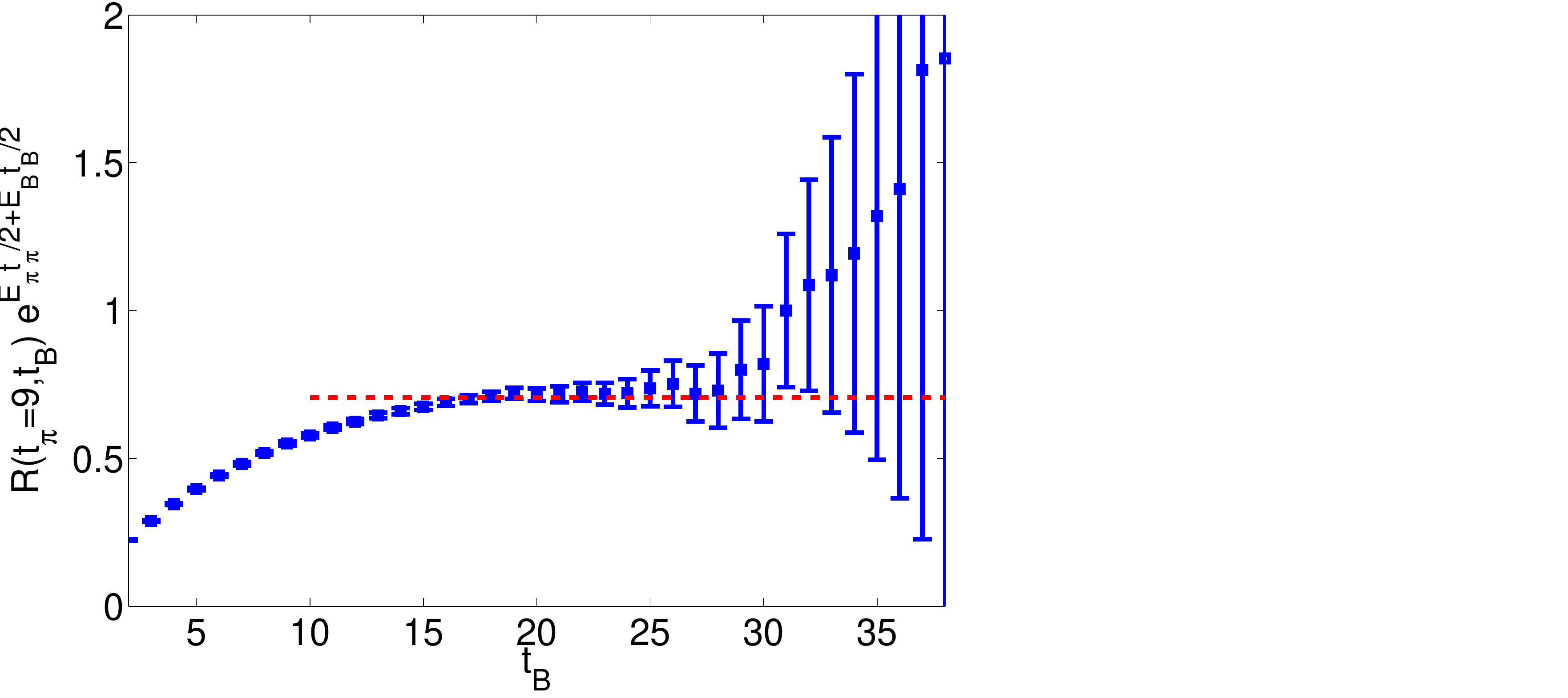}
\end{tabular}
\end{center}
\caption{Left (Right): $t_\pi$ ($t_{\rm B}$) dependence of the ratio in 
eq.~\protect \eqref{ratio} at fixed $t_{\rm B}$ ($t_\pi$) for ensemble F6 
(see Table ~\protect \ref{quarkCLS} for detailed parameters)
and smeared wave functions for both the B and the $\pi$. The red curve is the expected finite-$T$ behaviour fitted according to eq.~\protect \eqref{ratiofit}.
}
\label{plateaux}
\end{figure}
 The second is a weighted average over the plateau region, which has a relatively small extent due to the combined presence of excited states and large finite-$T$ effects. The quality of our plateaux is shown in Fig.~\ref{plateaux}, where the 
$t_\pi$ ($t_{\rm B}$) dependence at fixed $t_{\rm B}$ ($t_\pi$) is plotted for one of our 
lattices (F6, see Table~\ref{quarkCLS}). We note that the 
smearing in the pion sector is very efficient and a plateau sets in at early 
$t_\pi$, while in the B sector excited state effects are quite large and the
plateau only starts when the noise is already quite sizable. This is quite
different from the computation of the B-meson energy and decay constant from
two-point functions~\cite{Bernardoni:2012}.
\\
Both our methods to extract the form factors give consistent results.
All results presented in this work were obtained with the two-parameter fit method. 
We have also checked that results obtained at different time slices $t_{\rm B}$ (inside the plateau region), or with a different fit range in $t_\pi$ are compatible. The errors were computed using the methods described in \cite{Schaefer:2010hu}. \\

\begin{figure}
\begin{center}
\begin{tabular}{cc}
\includegraphics[width=7cm]{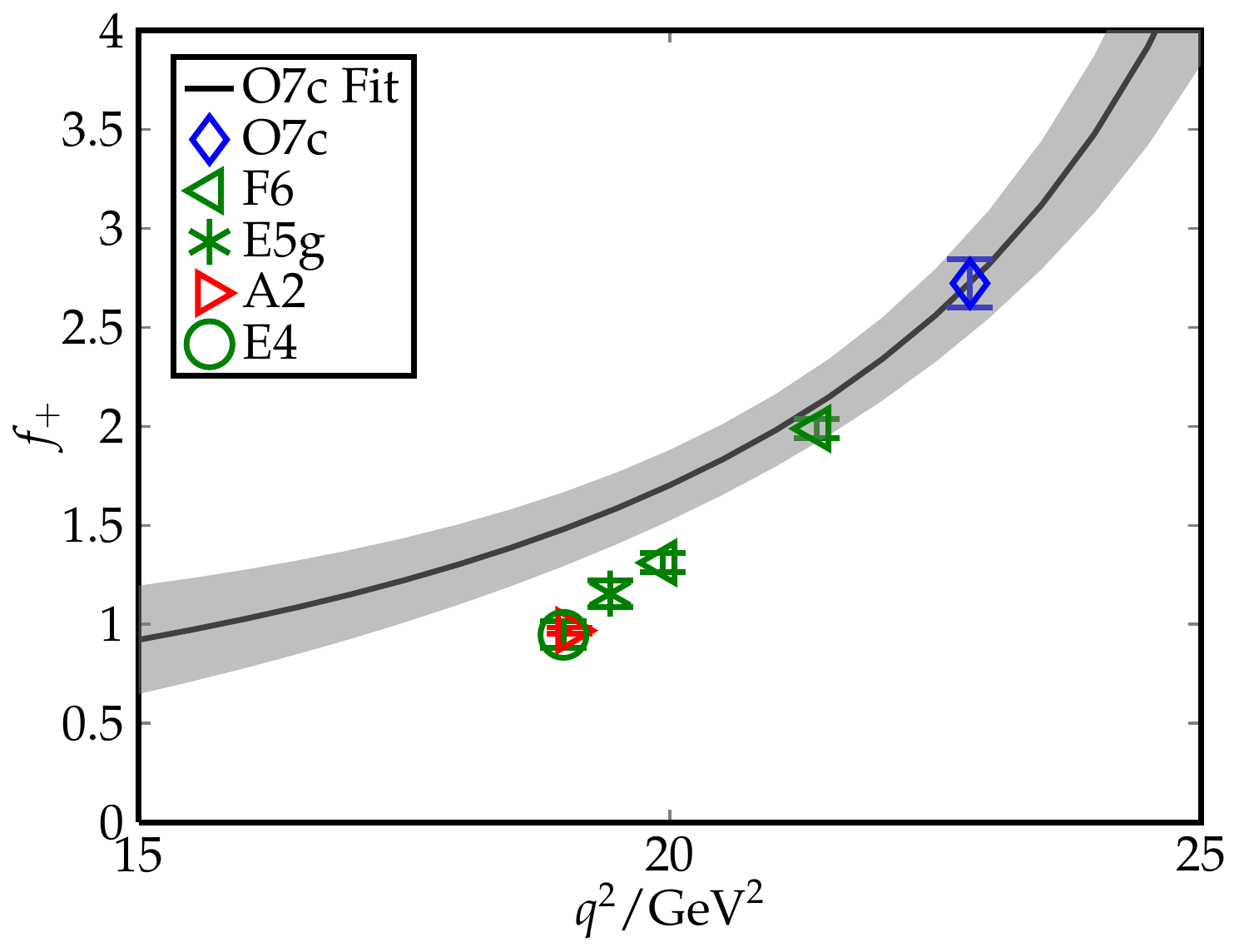}
&
\includegraphics[width=7cm]{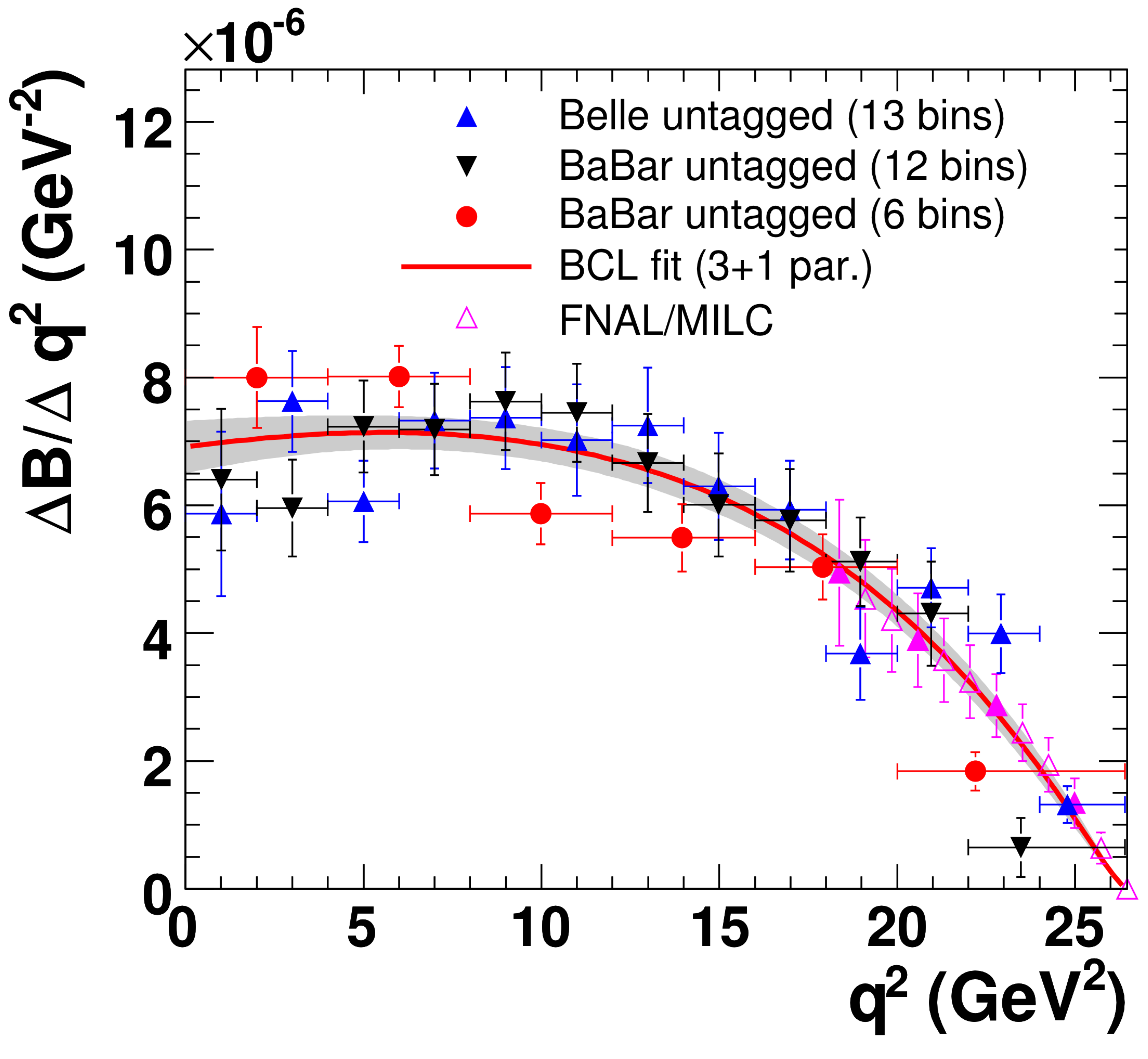}
\end{tabular}
\end{center}
\caption{Fit of the $q^2$-dependence of the form factor using the
  parametrization proposed in \protect \cite{Bourrely:2008za}.  The left panel shows a
   fit using only the
  computation from O7 and the experimental data available. The figure from
  \cite{Beringer:1900zz} in the right panel illustrates the experimental situation. Note that points at lower $q^2$ are more affected by
  $O(a)$ and $1/m_b$ effects.}
\label{results}
\end{figure}

\section{Conclusions and Outlook}

The results collected so far do not yet allow an extrapolation to the continuum 
limit or to the physical pion mass. 
Moreover, the vector current is not yet $O(a)$-improved
beyond tree-level of perturbation theory  
and we work in the static approximation of HQET. Therefore, cutoff effects of
order $aE_\pi\,\alpha_s$ 
and truncation errors of order $O(E_\pi/m_{\rm B})$ are present. 
We are currently working to include in our computation all $O(a)$ and $1/m_b$ terms and to perform the matching with QCD at non-perturbative accuracy.\\
The encouraging outcome of our present study is that we are able to extract a
signal for the form factor on our ensembles with a precision of around $5-10\%$. Using a parametrization of the $q^2$-dependence of the form factor \cite{Bourrely:2008za},
we have fitted the available experimental data together with our determination 
of $f_+(q^2)$ at the smallest available $p_\pi$ (largest $q^2$) from 
just one of our ensembles (O7 which is closest to the physical point), 
This parametrization is based on the so called z-expansion and very general principles, like analyticity, unitarity, and asymptotic freedom. In this fit $|V_{\rm ub}|$ appears as a free parameter and can be determined with a precision of around $15\%$.\\
With future more precise experimental data and including the $O(1/m_b)$ terms in our lattice analysis, 
we hope to reduce the uncertainty on $|V_{\rm ub}|$ to the $5-10\%$ level. 
Of course, eventually we will need to consider $N_{\rm f}>2$, but we would first
like to verify with $N_{\rm f}=2$ simulations that the extrapolation to the 
continuum limit can be carried out within our framework of non-perturbatively 
renormalized HQET.

\bibliography{lattice}
\bibliographystyle{plain}
\end{document}